\documentclass[twocolumn,showpacs,superscriptaddress,preprintnumbers,amsmath,amssymb,prc]{revtex4-2}

\usepackage{graphicx}
\usepackage{dcolumn}
\usepackage{bm}
\usepackage{float}
\usepackage{color}
\usepackage{CJK}
\usepackage{ulem}
\usepackage{array}
\usepackage{makecell}

\usepackage[colorlinks,linkcolor=blue,urlcolor=blue,citecolor=blue]{hyperref}

\usepackage{tikz,xcolor,hyperref}

\definecolor{lime}{HTML}{A6CE39}
\DeclareRobustCommand{\orcidicon}{
	\begin{tikzpicture}
	\draw[lime, fill=lime] (0,0) 
	circle [radius=0.16] 
	node[white] {{\fontfamily{qag}\selectfont \tiny ID}};
	\draw[white, fill=white] (-0.0625,0.095) 
	circle [radius=0.007];
	\end{tikzpicture}
	\hspace{-2mm}
}
\foreach \x in {A, ..., Z}{%
	\expandafter\xdef\csname orcid\x\endcsname{\noexpand\href{https://orcid.org/\csname orcidauthor\x\endcsname}{\noexpand\orcidicon}}
}
\foreach \x in {A, ..., Z}{%
	\expandafter\xdef\csname orcid\x\endcsname{\noexpand\href{https://orcid.org/\csname orcidauthor\x\endcsname}{\noexpand\orcidicon}}
}

\begin{document}
\begin{CJK*}{UTF8}{gbsn}

\title{Influence of Cluster Configurations and Nucleon--Nucleon Scattering Cross-Section on Stopping Power in Heavy-Ion 
Collisions}

\author{S. Y. Yao (姚少雨)}
\affiliation{Key Laboratory of Nuclear Physics and Ion-beam Application (MOE), Institute of Modern Physics, Fudan University, Shanghai 200433, China}

\author{X. G. Deng (邓先概)\orcidA{}}
\email[]{xiangai\_deng@fudan.edu.cn}

\author{Y. G. Ma (马余刚)\orcidB{}}
\email[]{mayugang@fudan.edu.cn}
\affiliation{Key Laboratory of Nuclear Physics and Ion-beam Application (MOE), Institute of Modern Physics, Fudan University, Shanghai 200433, China}
\affiliation{Shanghai Research Center for Theoretical Nuclear Physics， NSFC and Fudan University, Shanghai 200438, China}

\date{\today}

\begin{abstract}

We investigate the impacts of nuclear $\alpha$-clustering structures and nucleon--nucleon cross-section on nuclear stopping power for ${}^{16}\text{O}$ + ${}^{40}\text{Ca}$ collisions below 300 MeV/nucleon  using an extended quantum molecular dynamics (EQMD) model. Our results show that the specific $\alpha$-clustering configurations of ${}^{16}\text{O}$--including chain, square, kite, and tetrahedron--have a significant effect on collision dynamics. Among them, the tightly bound tetrahedral structure exhibits the highest stopping power. 
Moreover, the repulsive Coulomb interaction is found to reduce the stopping power of protons in the Fermi-energy domain. At higher energies, the decreasing trend is influenced by both the nucleon--nucleon cross-section and the mean field.

\end{abstract}


\maketitle

\section{Introduction}
\label{introduction}

Heavy ion collisions (HICs) provide a venue for studying the properties of nuclear matter under different thermal and dense conditions, offering in sights into the equation of state (EOS) and transport properties of nuclear matter~\cite{EOS1,Li1998,Bertsch1988,Aichelin,Kuttan,Pang,He_SCPMA,XuJ}, particularly in intermediate- and high-energy collisions. Stopping power \cite{Lehaut2010,ZhangGQ,Song}, collective flow \cite{Ma_flow,Liu}, and fragmentation \cite{Ma_book,PPNP} are crucial indicators of nuclear system evolution during collisions, revealing mechanisms of energy dissipation and momentum transfer~\cite{Baran,Ono,PPNP,Wolter,PPNP2,DengXG}. These observables also provide essential constraints on transport models, especially for intermediate- and high-energy nuclear reactions~\cite{Wolter,LiBA}. A key observable for studying stopping power in HICs is the isotropy ratio \( R_E \), which measures the redistribution of energy from longitudinal to transverse direction during the collision. Another observable, introduced by the FOPI collaboration, is the transverse-to-longitudinal variance ratio (vartl), which reflects the relative anisotropy of the particle distribution, indicating the degree of stopping and the thermalization in nuclear systems~\cite{Reisdorf}. This observable has been widely used by experimental collaborations such as INDRA and ALADIN to analyze nuclear stopping at moderate energies~\cite{Lehaut2010,Lopez2014}. It is found that the stopping power observable is correlated with the amount of side flow~\cite{Reisdorf,Cozma} as well as the magnetic field and eccentricity \cite{Ind}. In the Fermi-energy domain, these observables are sensitive to the in-medium nucleon--nucleon cross-section in nuclear matter~\cite{Zhang2007,Lopez2014,Song,LiuHL}.

In recent years, there has been growing interest in the impact of initial geometric fluctuations in intermediate-energy heavy-ion reactions \cite{Ma_review,CCGUO2017,YTCao2023,Ye_NST} as well as in relativistic heavy-ion collisions \cite{PBo2014,ZhangS,SZhang2018,Sch,Wang,YYWang2024,LiF_1,Zhang_1,Jia_1,Rys,Jia_2,LiF_2,Xu_1,Zhang_2,Gia}. Meanwhile, experimental studies at the Relativistic Heavy Ion Collider (RHIC) have revealed structural deformation characteristics of atomic nuclei~\cite{CJZhang2024,Giacalone,Jia}. Therefore, it is essential to explore nuclear structure information using different observables in the moderate reaction energy ranges. On the one hand, this deepens our understanding on nuclear structure information, especially for clustering configurations (e.g., tetrahedron vs. chain of $^{16}$O) on nuclear matter properties. In some previous literature, it is shown that clustering at low density has significant effects on astrophysical nuclear equations of state \cite{JBN,Qin}.
On the other hand, it enables mutual validation, using different observables and/or different models. Therefore, in this work, we employ the extended quantum molecular dynamics (EQMD) model to study nuclear stopping power 
by simulating collisions between α-cluster structured ${}^{16}\text{O}$ projectiles and normally structured ${}^{40}\text{Ca}$ target nuclei. We explore different $\alpha$-clustering configurations of ${}^{16}\text{O}$, including chain, square, kite, and tetrahedral structures, to assess their impacts on energy dissipation and stopping observables. Additionally, we investigate the impact of various nucleon--nucleon cross-sections on stopping power observables,
examining how cross-section choices affect isotropy and collision dynamics. 

The paper is organized as follows: Section~\ref{modelandmethod} introduces the models and analysis methods in this work. Section~\ref{resultsSH} explores two factors: the role of the Coulomb potential in stopping and a comparison of stopping behaviors for different collision cross sections. Finally, the summary and conclusions are presented in Sec.~\ref{summary}.

\section{Model introductions and analysis method}
\label{modelandmethod}

\subsection{EQMD model}
\label{EQMDModel}
The extended quantum molecular dynamics model~\cite{MaruyamaNiitaIwamoto} is a sophisticated approach used in the simulation of heavy-ion collisions. It extends the traditional quantum molecular dynamics model by allowing greater flexibility in the description of nucleon wave packets, which makes it particularly useful for investigating the dynamics of nuclear matter at moderate energies. Unlike the standard QMD model~\cite{AichelinStocker,Aichelin}, which assumes constant widths for nucleon wave packets, EQMD introduces dynamically varying widths that allow for a more accurate representation of nucleon interactions and correlations. Some developments for the EQMD model are also reported in Refs.~\cite{Shi1,Shi2,Shi3}.

In EQMD, nucleons (both protons and neutrons) are represented as Gaussian wave packets. The nucleon's wave function is a Gaussian with time-dependent width. The total system's wave function can be described as a product of individual nucleon wave functions~\cite{MaruyamaNiitaIwamoto}:
\begin{equation}
\Psi = \prod_i \phi_i (\mathbf{r}_i),
\end{equation}
where \( \phi_i(\mathbf{r}_i) \) is the wave function of the \(i\)-th nucleon. The Gaussian form of each nucleon's wave function is given by:
\begin{equation}
\phi_i(\mathbf{r}_i) = \left( \frac{v_i + v_i^*}{2\pi} \right)^{3/4} 
\exp\left[ -\frac{v_i}{2} (\mathbf{r}_i - \mathbf{R}_i)^2 + \frac{i}{\hbar} \mathbf{P}_i \cdot \mathbf{r}_i \right],
\end{equation}
where \( R_i \) and \( P_i \) denote the nucleon's position and momentum centers in phase space. The width \( v_i \) is a complex quantity, given by $v_i = \frac{1}{\lambda_i} + i\delta_i$, with \( \lambda_i \) being the real part related to the spatial extension of the wave packet, and \( \delta_i \) the imaginary part describing its phase evolution. The real part \( \lambda_i \) must be positive to ensure a valid physical nucleon density.

The probability or density distribution for each nucleon has a Gaussian form, expressed as:
\begin{equation}
\rho_i(\mathbf{r}) = \frac{1}{(\pi \lambda_i)^{3/2}} \exp\left[ - \frac{(\mathbf{r} - \mathbf{R}_i)^2}{\lambda_i} \right].
\end{equation}

This formula shows that the spatial distribution of a nucleon depends on its position center \( R_i \) and the dynamically evolving width \( \lambda_i \). Allowing \( \lambda_i \) to change over time makes EQMD a flexible and adaptable model for capturing nuclear dynamics during collisions.

The Hamiltonian of the system, which governs its dynamics, can be written as:
\begin{equation}
\begin{split}
H & =  \langle \Psi | \sum_i \left( -\frac{\hbar^2}{2m} \nabla_i^2 - \hat{T}_{\text{c.m.}} \right) + \hat{H}_{\text{int}} | \Psi \rangle 
\\
&= \sum_i \left( \frac{\mathbf{P}_i^2}{2m} + \frac{3\hbar^2}{2} \frac{1 + \lambda_i^2 \delta_i^2}{2m\lambda_i} - T_{\text{c.m.}} \right) + H_{\text{int}},
\end{split}
\end{equation}
where \( \hat{T}_{\text{c.m.}} \) is the zero-point kinetic energy due to the center-of-mass motion, and \( \hat{H}_{\text{int}} \) represents the potential interactions between nucleons. 

The effective interaction $H_{\text{int}}$ is described by
\begin{equation}
H_{\text{int}} = H_{\text{Sky.}} + H_{\text{Coul.}} + H_{\text{Symm.}} + H_{\text{Pauli}},
\end{equation}
where \( H_{\text{Sky.}} \), \( H_{\text{Coul.}} \), \( H_{\text{Symm.}} \), and \( H_{\text{Pauli}} \) represent the Skyrme, Coulomb, symmetry, and Pauli potentials, respectively.
The Skyrme potential, consisting of two-body and three-body terms, is:
\begin{equation}
H_{\text{Sky.}} = \frac{\alpha}{2\rho_0} \int \rho^2(\mathbf{r}) \, d\mathbf{r} + \frac{\beta}{(\gamma+1)\rho_0^\gamma} \int \rho^{\gamma+1}(\mathbf{r}) \, d\mathbf{r},
\label{EqSkyrme}
\end{equation}
where \( \alpha \), \( \beta \), and \( \gamma \) are parameters associated with the incompressibility of the nuclear equation of state. In this work, we use \( \alpha = -124.3\,\text{MeV},\ \beta = 70.5\,\text{MeV},\ \gamma = 2 \). At saturation density of nuclear matter $\rho_0 = 0.16~\mathrm{fm}^{-3}$, binding energy per nucleon E/A and the incompressibility $K$ are -16.5 MeV and 380 MeV, respectively. The symmetry potential, simplified, is
\begin{equation}
H_{\text{Symm.}} = \frac{c_s}{2\rho_0} \int \left( \rho_p - \rho_n \right)^2 \, d\mathbf{r},
\end{equation}
where \( \rho_p \) and \( \rho_n \) denote proton and neutron densities, respectively. The symmetry energy coefficient is taken as \( c_s = 25\,\text{MeV} \). The Coulomb potential considers only the direct term and is given by
\begin{equation}
H_{\text{Coul.}} = \frac{\alpha_{\text{0}}}{2} \int d\mathbf{r}\, d\mathbf{r'} \, \frac{\rho_p(\mathbf{r}) \rho_p(\mathbf{r'})}{|\mathbf{r} - \mathbf{r'}|}.
\end{equation}
with $\alpha_{\text{0}}$ being the fine structure constant.
Finally, the Pauli potential reflects the Fermi nature of nucleons and is
\begin{equation}
H_{\text{Pauli}} = \frac{c_P}{2} \sum_i \left( f_i - f_0 \right)^\mu \theta(f_i - f_0).
\end{equation}
Here, \( f_i \) is the overlap of the \( i \)-th nucleon with other nucleons having the same spin and isospin, i.e., \( f_i \equiv \sum_j \delta(S_i, S_j)\delta(I_i, I_j)|\langle \phi_i | \phi_j \rangle|^2 \), and \( \theta \) is the unit step function. The parameters are set as \( c_P = 15\,\text{MeV} \), \( f_0 = 1.0 \), and \( \mu = 1.3 \). This Pauli potential inhibits the system to collapse into the Pauli-blocked state at low energy and gives the model the capability to describe $\alpha$-particle clustering \cite{HeMaCao}. This capability is very important for our calculation because it gives us the possibility to extract information about clustering configurations from the nuclear stopping. 

In the model, the equations of motion for the nucleons can be written as~\cite{MaruyamaNiitaIwamoto}
\begin{equation}
\begin{split}
\dot{\mathbf{R}}_i = \frac{\partial H}{\partial \mathbf{P}_i}, \quad \dot{\mathbf{P}}_i = -\frac{\partial H}{\partial \mathbf{R}_i},\\
\frac{3\hbar}{4} \dot{\lambda}_i = -\frac{\partial H}{\partial \delta_i}, \quad \frac{3\hbar}{4} \dot{\delta}_i = \frac{\partial H}{\partial \lambda_i}.
\end{split}
\end{equation}
These equations describe the time evolution of the nucleon position, momentum, and wave packet width parameters during the heavy-ion collision process. In the model, the cross section $\sigma_{NN}$ is energy-dependent nucleon--nucleon collision cross section parametrized as~\cite{MaruyamaNiitaIwamoto},
\begin{equation}
\sigma_{NN}=\frac{100}{1+\epsilon'/200~\rm{MeV}}~\rm{mb},  \epsilon'=p_{rel}^{2}/2m,
\label{EqSIGMANN}
\end{equation}
where $p_{rel}$ is relative momentum. The collision is assumed to be isotropic. The collision process in the simulation is mainly handled through three components: the reaction channel, the nucleon--nucleon cross-section, and the Pauli blocking mechanism.
Specifically, two nucleons are considered to undergo a collision during a time step $\Delta t$ if their minimum relative distance $d_{min}$ satisfies $d_{min} \leq d_0 = \sqrt{(\sigma_{NN}/\pi)}$. After satisfying the geometric condition, the Pauli blocking criterion is further applied to ensure the quantum statistical property of the system.

The phase space of nucleons is obtained initially from a random configuration. To get the energy-minimum state as a ground state, a frictional cooling method is used for the initialization process~\cite{MaruyamaNiitaIwamoto,HeMaCao}. The EQMD model can describe the ground state properties, such as binding energy, rms radius, and deformation, etc., quite well over a very wide mass range.

\subsection{GEMINI++}

The deexcitation of hot nuclear fragments produced in heavy-ion collisions is a critical process for understanding the final state of the system \cite{Ma-2002}. To accurately model this process, we employ the GEMINI++ code~\cite{RJCharity2008, Charity2010}, which is an improved version of the GEMINI model developed by Charity~\cite{Charity1988}. GEMINI++ is a Monte Carlo simulation tool designed to handle various decay modes, including light-particle evaporation and symmetric fission~\cite{Charity1988, Charity2010}. Following the initial evolution of nuclear interactions simulated by the EQMD model, hot fragments are generated, each characterized by its charge number \(Z\), mass number \(A\), and excitation energy \(E^*\). The EQMD code halts when the excitation energy of the heaviest fragments falls below a specified threshold \(E_{\text{stop}}\).

If the excitation energy is greater than zero, the sequential decays of the prefragments are simulated using the GEMINI++ code. The decay width for the emission of a light particle \((Z_i, A_i)\) with spin \(S_i\) from a system \((Z_0, A_0)\) with excitation energy \(E^*\) and spin \(S_{\text{CN}}\), leaving a residual system \((Z_d, A_d)\) with spin \(S_d\), is calculated using the Hauser-Feshbach formalism~\cite{Hauser1952}:
\begin{equation}
\Gamma_i^{\text{HF}} = \frac{1}{2\pi \rho_0} \int d\epsilon \sum_{S_d=0}^{+\infty} \sum_{J = |S_{\text{CN}} - S_d|}^{S_{\text{CN}} + S_d} \sum_{l = |J - S_i|}^{J + S_i} T_l(\epsilon) \rho(U, S_d),
\end{equation}
where \( \epsilon \) is the kinetic energy of the emitted particle, \( J \) is the total angular momentum, and \( T_l(\epsilon) \) is the transmission coefficient or barrier penetration factor. The thermal excitation energy \(U\) is given by
\begin{equation}
U = E^* - E_{\text{rot}} - B_i - \epsilon,
\end{equation}
where \(E_{\text{rot}}\) is the rotational energy of the ground-state configuration, and \(B_i\) is the separation energy calculated from nuclear mass tables.

To model the nuclear level densities, we adopt the standard Fermi-gas model:
\begin{equation}
\rho(E^*, J) \propto (2J + 1) \frac{\exp\left[ 2\sqrt{a(E^* - E_1)} \right]}{a^{1/4} (E^* - E_1)^{5/4}},
\end{equation}
where \(a\) is the level-density parameter and \(E_1\) is the energy backshift, both of which are treated as free parameters that account for shell effects.

The level-density parameter \(a\) is expressed as \cite{Egidy2005}:
\begin{equation}
a = [0.127 + 0.00498 (S(Z,N) - \delta^{'}) - 0.0000895A] A,
\end{equation}
where 
\begin{equation}
\delta^{'} =
\begin{cases}
+0.5 P_d, & \text{even-even nuclei} \\
0, & \text{odd-A nuclei} \\
-0.5 P_d, & \text{odd-odd nuclei.}
\end{cases}
\end{equation}

The deuteron pairing energy \( P_d \) is computed from experimental mass tables as:
\begin{align}
P_d = \frac{1}{2} (-1)^Z \big[ 
&-M_\text{exp}(A+2,Z+1) + 2M_\text{exp}(A,Z) \notag \\
&- M_\text{exp}(A-2,Z-1) \big],
\end{align}
where \( M_\text{exp} \) represents the experimental mass of the nucleus. The derivative of the shell correction energy \( S(Z,N) \) with respect to mass number \( A \) is approximated using finite differences:
\begin{equation}
\frac{dS(Z,N)}{dA} = \frac{S(Z+1,N+1) - S(Z-1,N-1)}{4}.
\end{equation}
 Using these formulations, GEMINI++ provides a comprehensive framework for simulating the de-excitation of nuclear fragments, leading to insights into the production mechanisms of heavy fragments in nuclear reactions. More details can be found in Ref.~\cite{Egidy2005}.

\subsection{Nuclear stopping}

To describe nuclear stopping power, we use the ratio of beam energy transferred to the parallel direction, \( R_E \), which is expressed as Ref.~\cite{Lehaut2010}:
\begin{equation}
R_E = \frac{\sum E_\perp}{2 \sum E_\parallel},
\end{equation}
where \( E_\perp \) (\( E_\parallel \)) is the transverse (parallel) kinetic energy of the fragments in the center-of-mass system. This ratio effectively reflects the isotropic characteristics of nuclear reactions.
Specifically, if \( R_E < 1 \), it indicates partial transparency; when \( R_E = 1 \), it represents full stopping; and if \( R_E > 1 \), it indicates super-stopping.

Furthermore, vartl, first proposed by the FOPI Collaboration~\cite{Reisdorf}, is defined as the ratio of variances of transverse and parallel rapidity distributions. The variance is calculated as the average over multiple events. Its expression is

\begin{equation}
\text{\(vartl\)}=\left\langle\frac{\sigma(y_\perp)}{\sigma(y_\parallel)}\right\rangle=\left\langle\frac{\sqrt{\frac{1}{N} \sum_{i=1}^N (y_{\perp, i} - \langle y_\perp \rangle)^2}}
{\sqrt{\frac{1}{N} \sum_{i=1}^N (y_{\parallel, i} - \langle y_\parallel \rangle)^2}}
\right\rangle
\label{Eqvartl},
\end{equation}
where \( \sigma(y_\perp) \) and \( \sigma(y_\parallel) \) represent the variances of transverse and parallel rapidity distributions, respectively. If the nuclear collision is fully stopped, the rapidity distribution will be isotropic.
In the calculations of \( R_E \) and \( \text{vartl} \), we did not strictly follow the formulas on an event-by-event basis. It was observed that computing \( R_E \) for each individual event led to fluctuations in the results, especially at lower energy ranges with fewer free nucleons emitted. To avoid this issue, we averaged \( R_E \) over 100 events. For Figs.~\ref{fig:fig5} and \ref{fig:fig6}, since fragment information is not available, we calculated \( R_E \) in nucleon space.

\section{Results and discussion}
\label{resultsSH}

In the EQMD model, the first essential step of collision simulation is to construct a physically reasonable initial nucleus. This initialization is typically achieved through frictional cooling, in which a dissipation term is introduced during the dynamical evolution to gradually reduce the system's kinetic energy, eventually reaching a stable state with minimal internal energy. After the cooling process, the resulting \(^{16}\mathrm{O}\) nucleus exhibits an average binding energy of \(-7.91~\mathrm{MeV}\) per nucleon (chain: -7.82 MeV, kite: -7.79 MeV, square: -7.92 MeV, and tetrahedron: -7.82 MeV), which is very close to the experimental ground-state value of \(-7.98~\mathrm{MeV}\). Here we did not specifies any of these configurations as the ground state of $^{16}$O since their binding energies are close to each other, in particular, we aim to compare different $\alpha$-clustering configuration effects in the present work.  We also note that the binding energies obtained here differ from those reported in the previous literature~\cite{WBHe_2016}, mainly because the parameter $C_{p}$ used in the present work is 15 MeV, whereas a value of 20 MeV was adopted in that reference. The Pauli potential plays a crucial role in this process. By suppressing the collapse of the system into Pauli-blocked states at low energies, it enables the model to naturally form  \(\alpha\) clusters~\cite{HeMaCao,Huang_1,Guo_1,Li_1,Ma_1}. For \(^{16}\mathrm{O}\), different structural configurations naturally emerge during cooling process and then can be classified as chain, square, kite, or tetrahedral shapes. In contrast, with the increase of the nucleon number of atomic nuclei, the nucleus, e.g., \(^{40}\mathrm{Ca}\) in this work, is more random without specific structural patterns and is therefore treated as a spherical nucleus.

Through the EQMD model, we studied the effects of $\alpha$-clustering structure of \(^{16}\mathrm{O}\), including chain, square, kite, and tetrahedron on the nuclear stopping. Since \(^{16}\mathrm{O}\) rotates uniformly around its central axis during the collision, we can roughly order its nuclear density distribution in the beam direction as \( \text{chain} < \text{kite} < \text{square} < \text{tetrahedron} \). After the collision, we selected fragments with excitation energy greater than 0 and mass number greater than 4 to decay using the GEMINI++ model.


\begin{figure}[htb]
\centering
\setlength{\abovecaptionskip}{0pt}
\setlength{\belowcaptionskip}{8pt}
\includegraphics[scale=0.5]{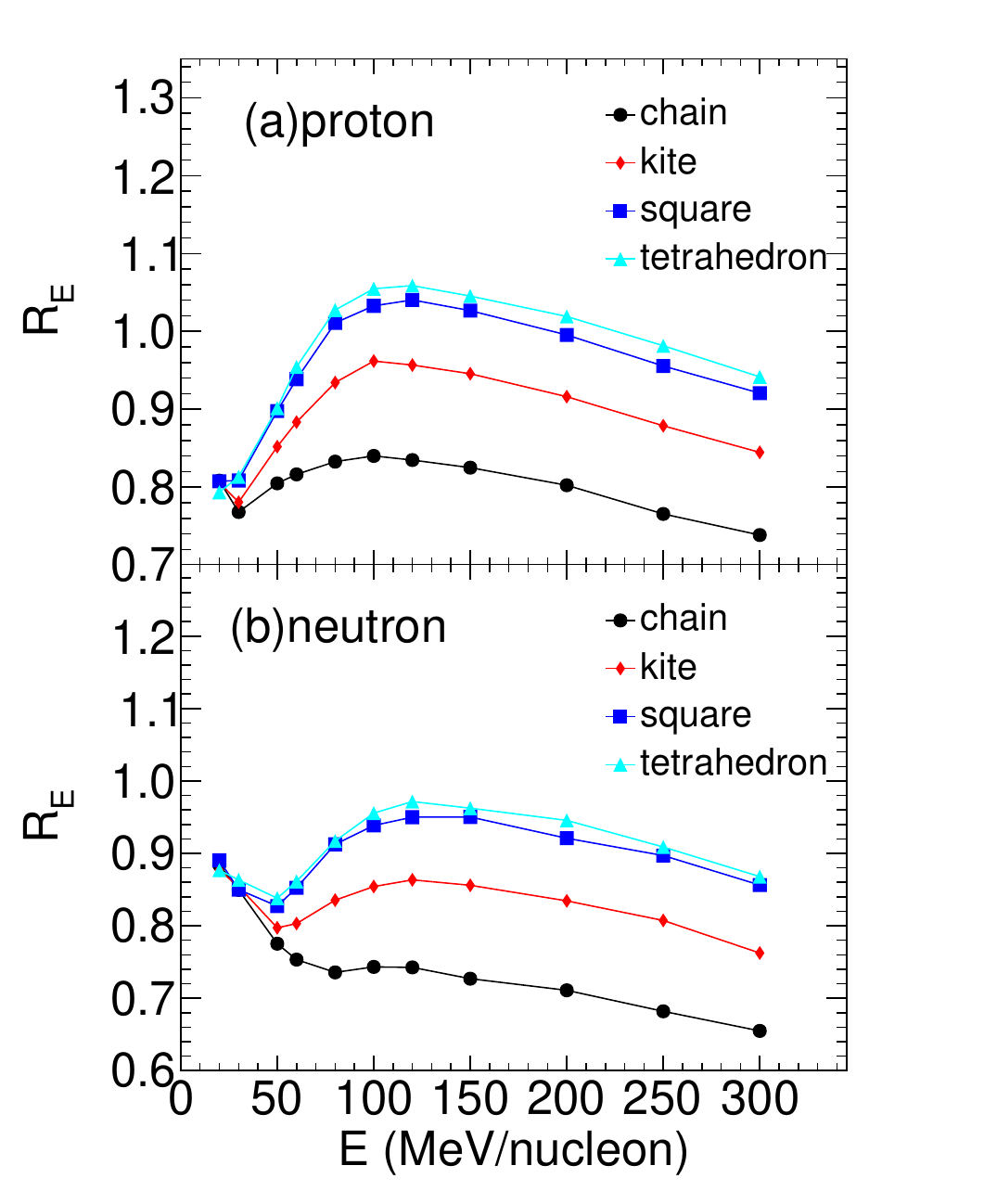}
\caption{The stopping $R_E$ of proton and neutron as a function of beam energy.} 
\label{fig:fig1}
\end{figure}

\begin{figure}[htb]
\setlength{\abovecaptionskip}{0pt}
\setlength{\belowcaptionskip}{8pt}
\includegraphics[scale=0.5]{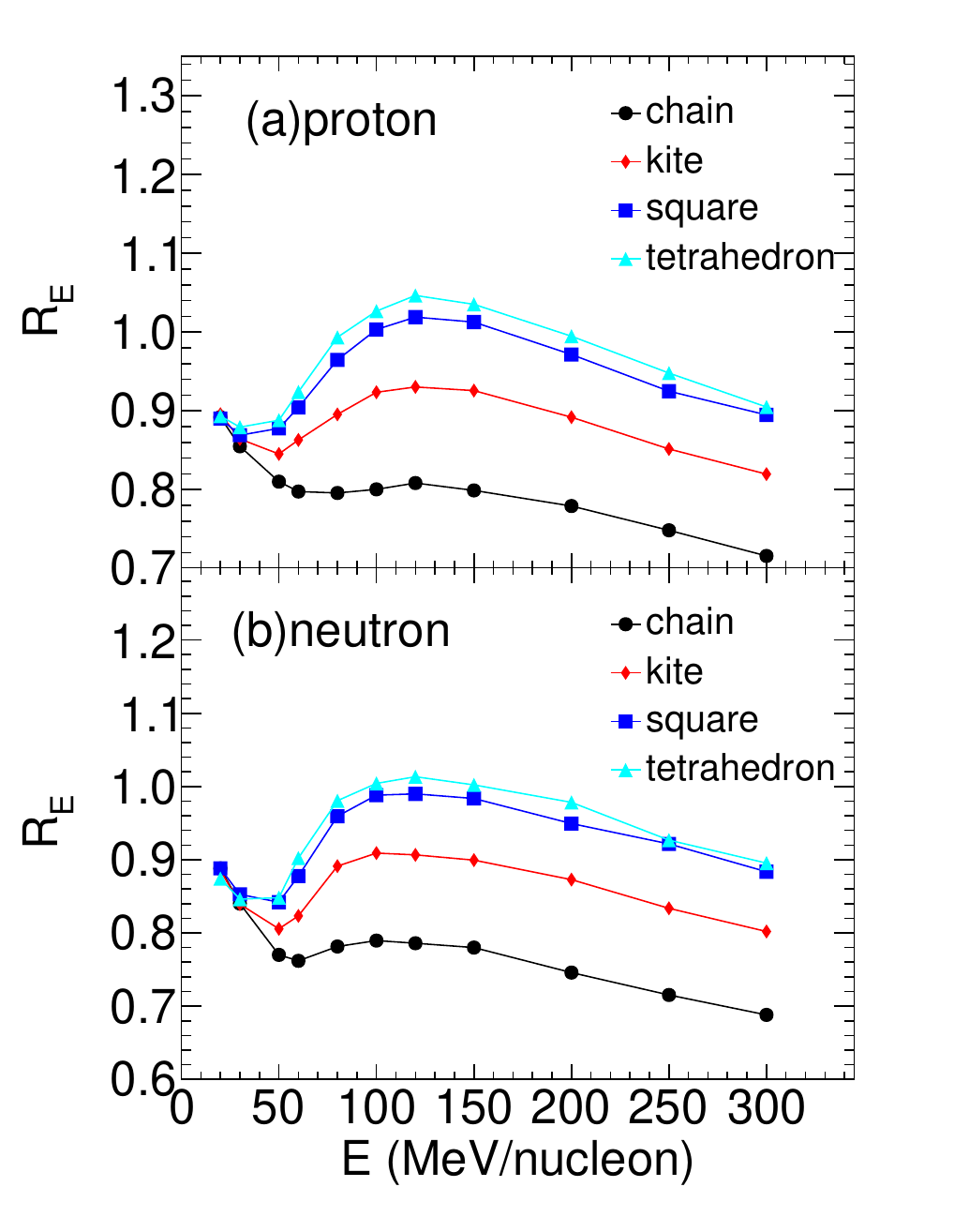}
\caption{The same as Fig.~\ref{fig:fig1}, but without Coulomb interaction.} 
\label{fig:fig2}
\end{figure}

As displayed in Fig.~\ref{fig:fig1}, one can find that the stopping power $R_E$ of neutrons and protons is sensitive to nuclear clustering structures. For these four configurations, the stopping power follows a hierarchy tetrahedron $>$ square $>$ kite $>$ chain. From the geometric arrangement, we know that the tetrahedron configuration is more compact than the chain configuration. The more compact the structure, the higher the central density and thus, the greater the nuclear stopping power. This also implies that there is a corresponding order of compactness among these four configurations: tetrahedron $>$ square $>$ kite $>$ chain. 
 Interestingly, it is found that this hierarchy of nuclear stopping is the opposite of the order of fragment fluctuations \cite{SunKJ,Deng,Ko}, which are represented by double ratios of light nuclei, namely proton, deuteron, triton or $^4$He \cite{YTCao2023}, indicating an inverse relationship between nuclear stopping and fragment fluctuations.
In Fig.~\ref{fig:fig1}, we also note that in the 20-50 MeV/nucleon range, the neutron $R_E$ decreases with increasing energy. This is consistent with experimental results~\cite{Lehaut2010}. In the reference, the cause responds to the strong decrease of the fusion cross section in this bombarding energy range~\cite{Lehaut2010,INDRA2006}. Here, we also consider an intrinsic reason: at this low energy region, the mean field and Pauli blocking dominate, and the increase in energy leads to an increase in the repulsive potential. When the compressional impact reaches its maximum, it tends to expand in the direction of the beam. This reduces the transfer of transverse momentum into longitudinal one, which means a smaller $R_E$. 

However, for protons, no decreasing trend is observed in the 20-50 MeV/nucleon range even though the neutron-proton symmetric projectile and target system is used in this work. The main reason for this difference is due to Coulomb interactions. To verify this, in Fig.~\ref{fig:fig2}, we turn off the Coulomb interaction, and it can be seen that both the proton and neutron $R_E$ exhibit a decreasing behavior in the 20-50 MeV/nucleon range. There is also no longer a distinction in the magnitude of $R_E$.
\begin{figure}[htb]
\setlength{\abovecaptionskip}{0pt}
\setlength{\belowcaptionskip}{8pt}
\includegraphics[scale=0.5]{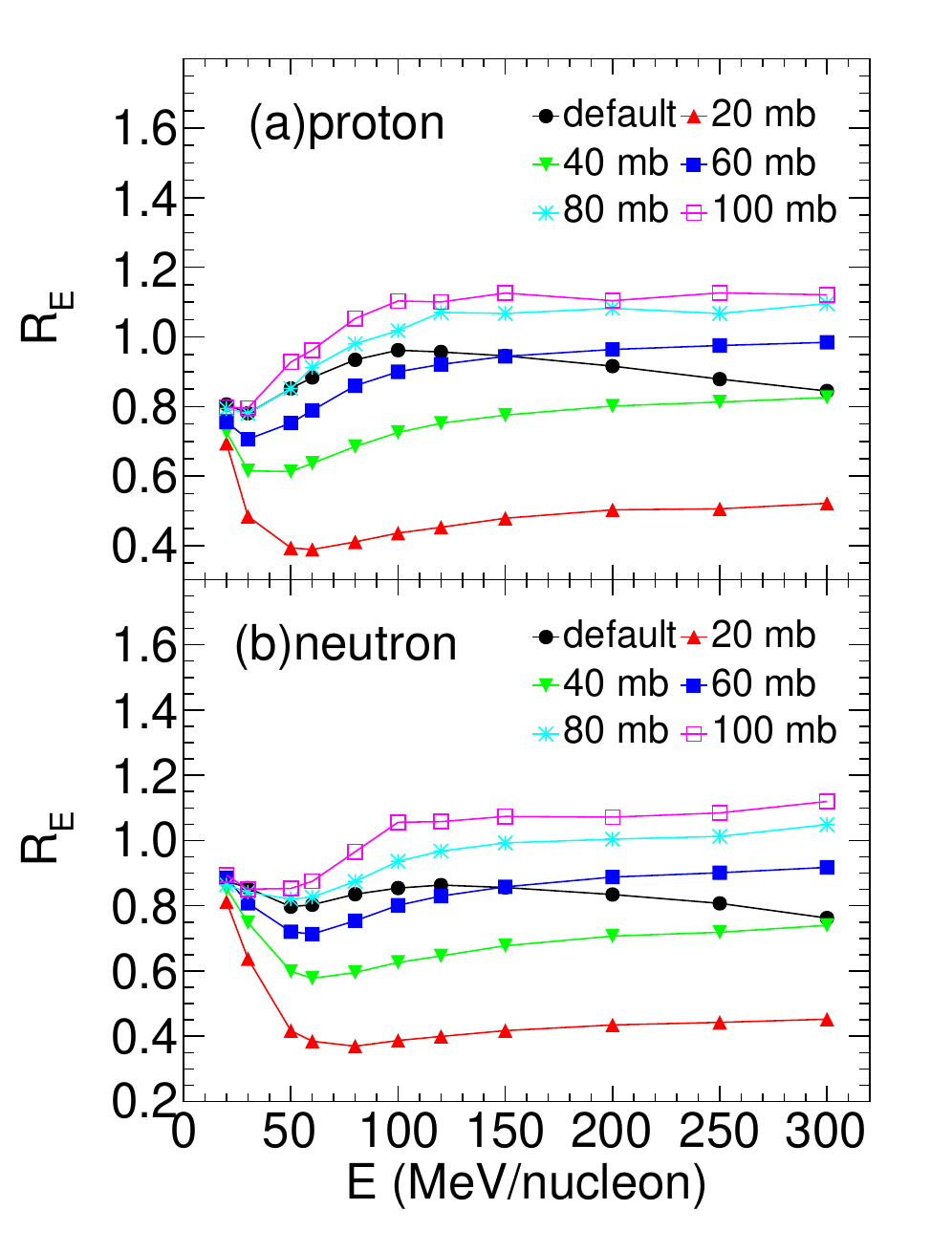}
\caption{ The $R_E$ as a function of beam energy for protons and neutrons at different collision cross sections using the kite configuration.}    
\label{fig:fig3}
\end{figure}

\begin{figure}[htb]
\setlength{\abovecaptionskip}{0pt}
\setlength{\belowcaptionskip}{8pt}
\includegraphics[scale=0.4]{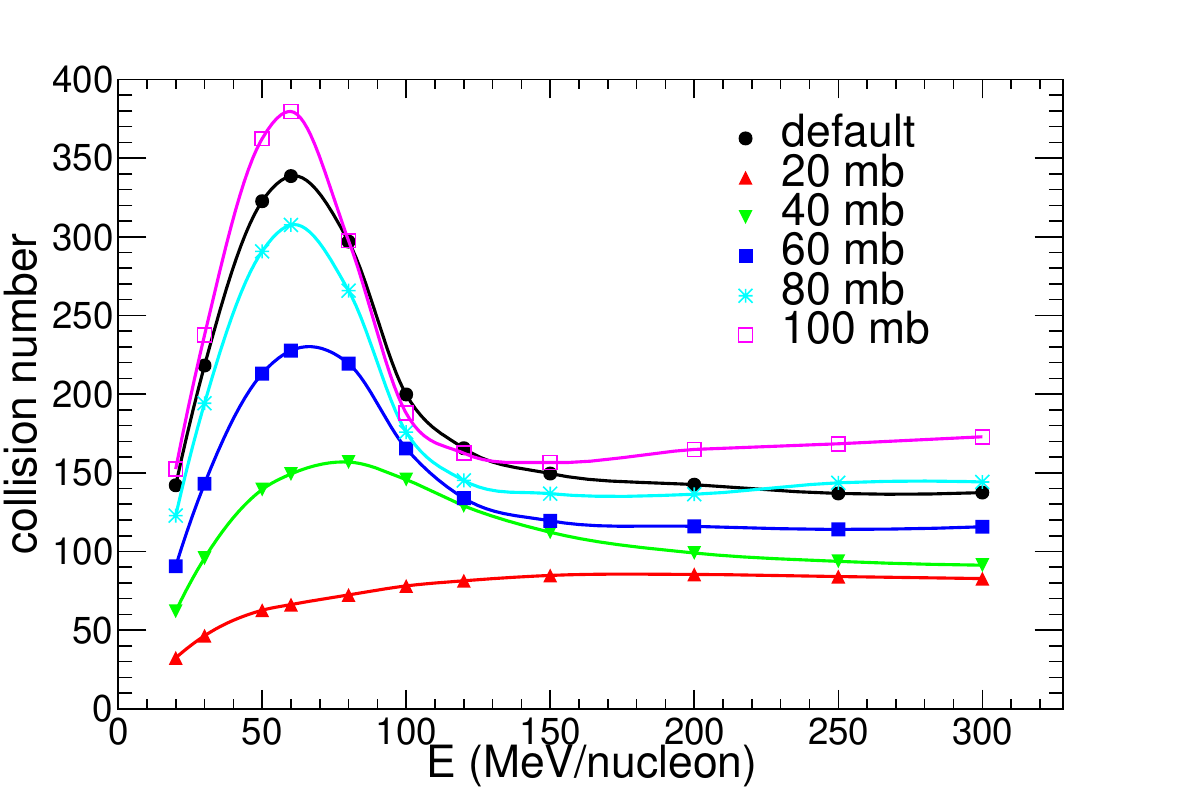}
\caption{The number of collisions as a function of beam energy for the kite configuration under different collision cross sections.}
\label{fig:fig4}
\end{figure}

\begin{figure}[htb]
\setlength{\abovecaptionskip}{0pt}
\setlength{\belowcaptionskip}{8pt}
\includegraphics[scale=0.5]{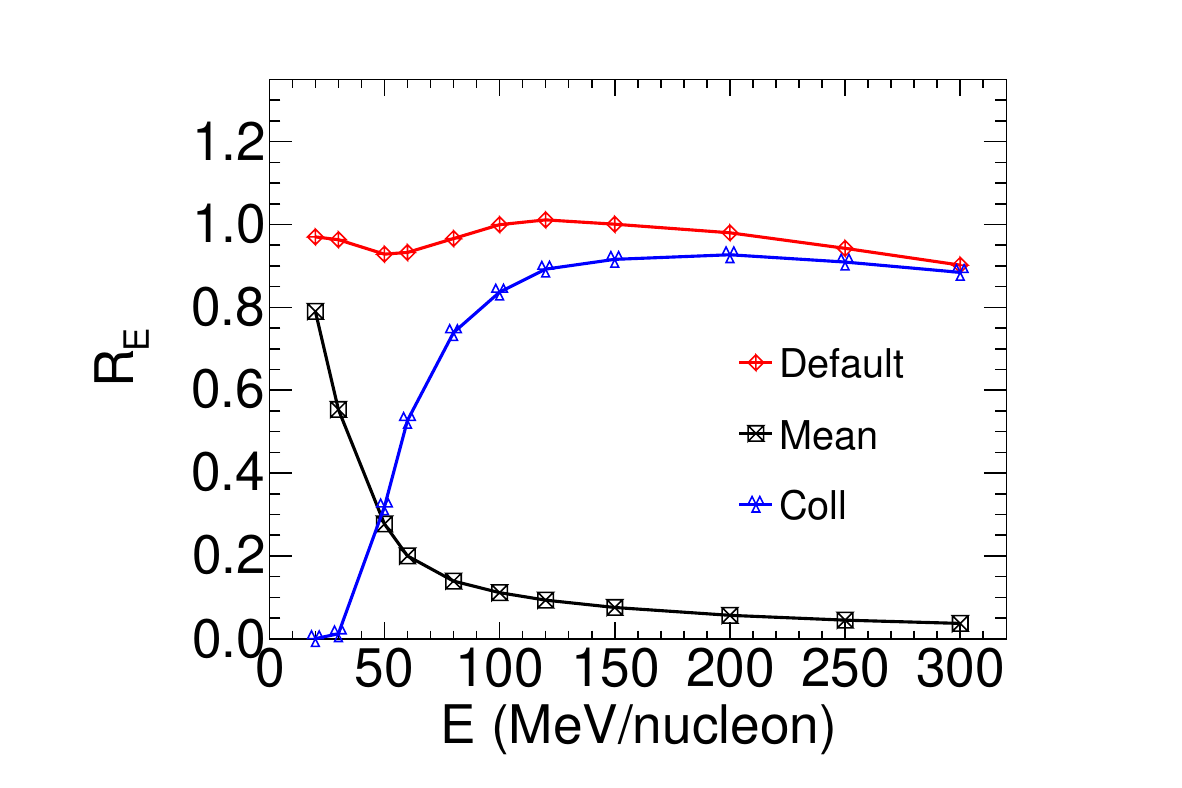}
\caption{ The $R_E$ as a function of beam energy for the tetrahedron configuration with different modes: Default （red curve),  Mean: only mean field (blue curve), and Coll: cascade (black curve). }
\label{fig:fig5}
\end{figure}

\begin{figure}[htb]
\setlength{\abovecaptionskip}{0pt}
\setlength{\belowcaptionskip}{8pt}
\includegraphics[scale=0.5]{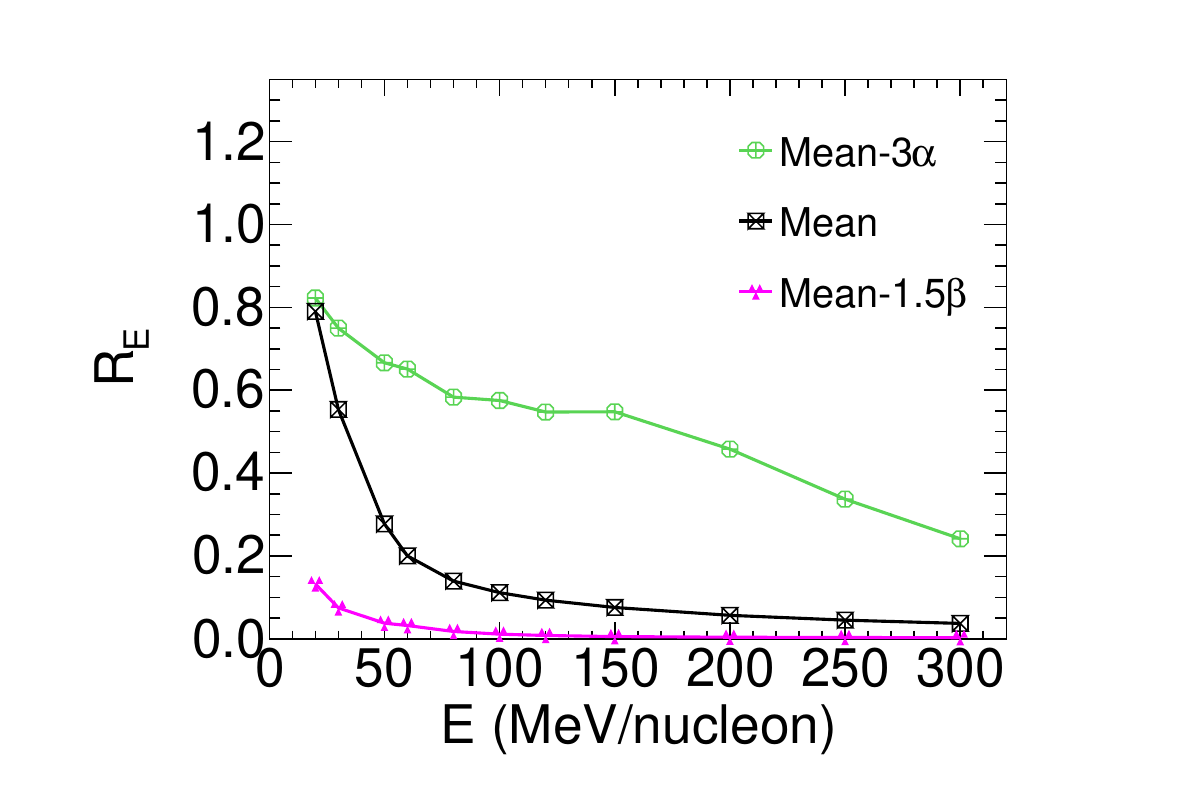}
\caption{The $R_E$ as a function of beam energy for the tetrahedron configuration with different modes:  Mean-3$\alpha $: only mean field with 3$\alpha$ (green curve);  Mean: only mean field (black curve), and  Mean-1.5$\beta$: only mean field with 1.5$\beta$ (pink curve).}
\label{fig:fig6}
\end{figure}

As the energy increases, two-body collisions increase, leading to an increase in $R_E$. Here, it can be seen that for the square and tetrahedron configurations, their values exceed 1. This indicates that in this energy regime, there is strong squeeze-out, with enhanced emission at polar angles of 90$^{\circ}$~\cite{Reisdorf}. Moreover, with a further increase in collision energy, $R_E$ decreases. The decrease in $R_E$ is due to the reduction of the cross section. In the model, the collision cross section is described by Eq.(~\ref{EqSIGMANN}). From the equation, it is clear that as the collision energy increases, the collision cross section decreases, which in turn leads to a decrease in $R_E$. To illustrate this, we performed simulations with different cross-sections. As shown in Fig.~\ref{fig:fig3}, at high collision energies, the stopping power increases with the increase in cross-section. Only with the default cross-section form does the stopping show a decreasing trend. Furthermore, we also present the dependence of the number of collisions on energy for different cross-sections, as shown in Fig.~\ref{fig:fig4}. For the same reaction energy, different collision cross sections lead to significant differences in the number of collisions. In general, a larger cross-section corresponds to a higher number of collisions. When the cross-section is large, the number of collisions first increases and then decreases. The maximum number of collisions indicates that the system has reached collision saturation. However, as the collision energy increases, the reaction time of the collision system decreases, leading to a reduction in the number of collisions. For small cross-sections, such as 20 mb, saturation is not reached. The energy point corresponding to the maximum value of $R_E$ does not coincide with the energy point where the number of collisions is maximal. This implies that there is no strong correlation between the number of collisions and the stopping power.

\begin{figure}[htb]
\setlength{\abovecaptionskip}{0pt}
\setlength{\belowcaptionskip}{8pt}
\includegraphics[scale=0.5]{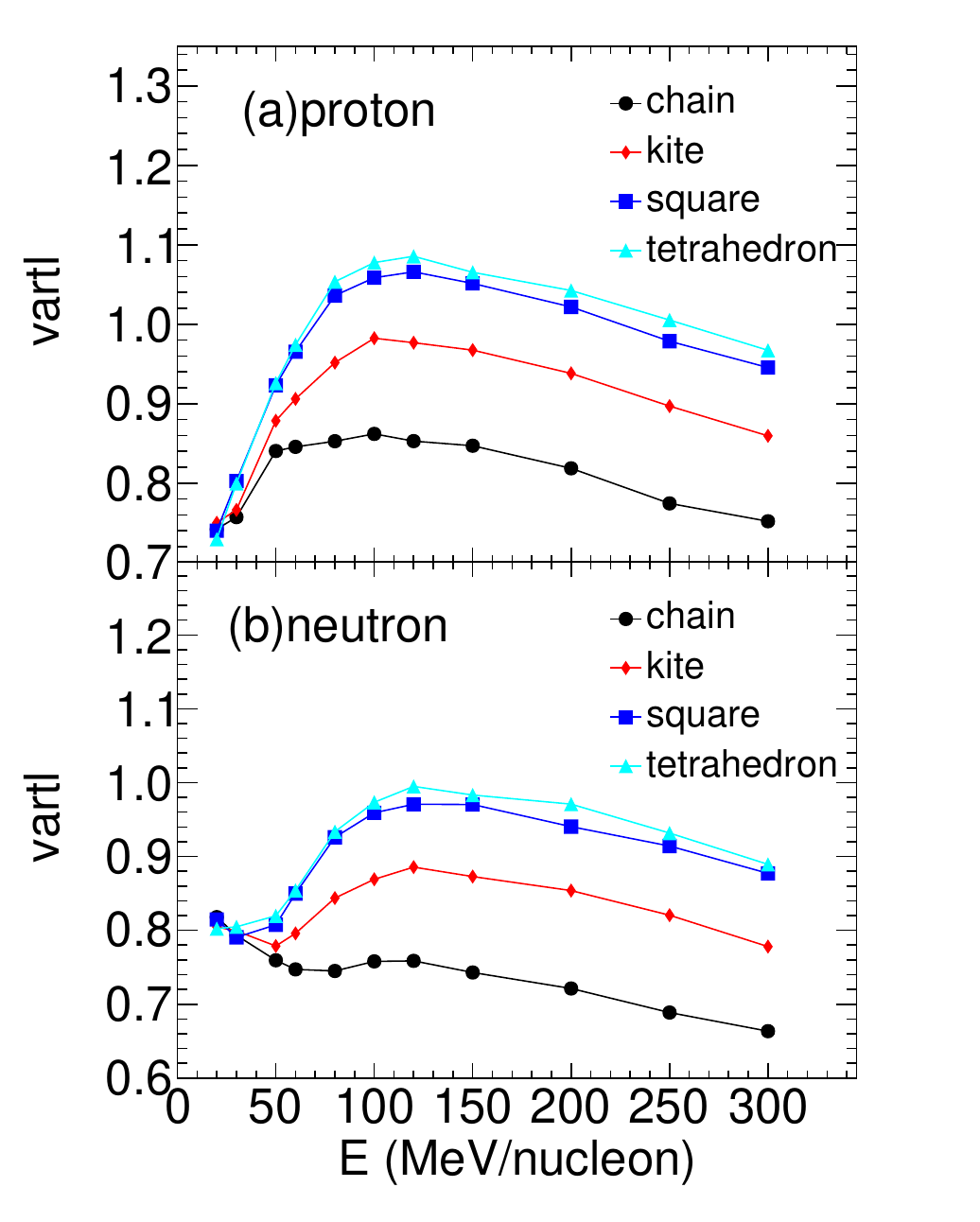}
\caption{The vartl of proton and neutron as a function of beam energy for different configurations.}
\label{fig:fig7}
\end{figure}

Additionally, we have tried to use some simplified modes as in Figs.~\ref{fig:fig5} and ~\ref{fig:fig6}. Note that since the fragment is unavailable in such modes, $R_E$ in Figs.~\ref{fig:fig5} and ~\ref{fig:fig6} are extracted from the nucleon space. Figure.~\ref{fig:fig5} presents the energy dependence of $R_E$ for three different modes: Default, Mean: only mean field and Coll: only cascade, i.e. mean-field is turned off and only collision term remains. It can be seen that when only the mean field is considered, $R_E$ monotonically decreases as energy increases. In the cascade mode, however, $R_E$ first increases with energy and then shows a decreasing trend. This suggests that the decrease of $R_E$ at high energies is not only due to the effect of the cross-section, but the mean field also plays a role. As in Fig.\ref{fig:fig6}, we enlarge the parameters $\alpha$ and $\beta$, which correspond to attractive and repulsive interactions in Eq.(\ref{EqSkyrme}), by multiplying them by 3.0 and 1.5, respectively. Figure.~\ref{fig:fig6} shows that when the attractive interaction increases, the stopping power increases. In contrast, when the repulsive interaction is increased, the stopping power decreases.

Finally, the another stopping observable is also presented, namely observable vartl as the Eq.~(\ref{Eqvartl}), which is also related to the $R_{E}$. As shown in Fig.~\ref{fig:fig7}, vartl is similar to $R_{E}$ as in Fig.~\ref{fig:fig1}. Essentially,  here vartl shows similar sensitivity to the different cluster configurations as $R_{E}$ does in Fig.~\ref{fig:fig1}, indicating the robust finding for the relationship of nuclear stopping versus the clustering configurations.

\section{Conclusions}
\label{summary}

This study used the EQMD model to investigate how nuclear cluster configurations affect the stopping power $R_{E}$ in central $^{16}$O+$^{40}$Ca collisions. The stopping power was found to be sensitive to the cluster structure, with the tetrahedral configuration exhibiting the highest stopping power, followed by the square, kite and chain configurations. More compact structures with higher central densities resulted in higher stopping power. Perhaps we cannot really determine the exact structure of a nucleus, but we may be able to probe the initial state of the nucleus {\color{red}\sout{-}} {\color{red}{---}}whether it is compact or diffuse {\color{red}\sout{-}} {\color{red}{---}} through the magnitude of the stopping power. Neutron stopping power decreased in the 20-50 MeV/nucleon range, while proton stopping power was affected by Coulomb interactions. At higher energies, two-body collisions initially increase the stopping power, but it decreases as the collision cross section decreases. The study also showed that the number of collisions correlates with the cross section, but not directly with the stopping power. Both mean field and cascade processes were found to influence the energy dependence of stopping power. Finally, changes in attractive and repulsive interactions affected stopping power, with attractive interactions increasing it and repulsive interactions decreasing it. 
These results improve our understanding of the role of nuclear cluster configurations and interactions on stopping power in heavy-ion collisions.

\begin{acknowledgments}
The authors thank Dr. Chen Zhong for maintaining the high-quality performance of Fudan HIRG supercomputing platform for nuclear physics. This work received partial support from the National Key R\&D Program of China Nos. 2023YFA1606701, the National Natural Science Foundation of China under Contract Nos. $12147101$, $12205049$, $12347149$, $11890714$, and $11925502$, the Strategic Priority Research Program of CAS under Grant No. XDB34000000, the Guangdong Major Project of Basic and Applied Basic Research No. 2020B0301030008, and the STCSM under Grant No. 23590780100.
\end{acknowledgments}

\end{CJK*}
\end{document}